\documentclass[12pt]{iopart}
\usepackage{siunitx}
\usepackage{graphicx}
\usepackage{hyperref}
\usepackage{subfigure}
\usepackage[style=ieee, citestyle=numeric-comp,backend=biber,sorting=none, doi=false,isbn=false,url=false, maxbibnames=2]{biblatex}
\begin{document}

\title[]{Comparison of linear and nonlinear methods for decoding selective attention to speech from ear-EEG recordings}

\author{Mike Thornton$^{(1)}$, Danilo Mandic$^{(2)}$, Tobias Reichenbach$^{(3)}$}

\address{$^{(1)}$Department of Computing, Imperial College London, SW7 2RH, U.K.\\
$^{(2)}$Department of Electrical and Electronic Engineering, Imperial College London, SW7 2RH, U.K.\\
$^{(3)}$Department Artificial Intelligence in Biomedical Engineering, Friedrich-Alexander-Universit\"at Erlangen-N\"urnberg, 91052 Erlangen, Germany}

\ead{tobias.j.reichenbach@fau.de}

\vspace{10pt}
\begin{indented}
\item[]July 2024
\end{indented}

\begin{abstract}
\textit{Objective.} Many people with hearing loss struggle to comprehend speech in crowded auditory scenes, even when they are using hearing aids. It has recently been demonstrated that the focus of a listener's selective attention to speech can be decoded from their electroencephalography (EEG) recordings, raising the prospect of smart EEG-steered hearing aids which restore speech comprehension in adverse acoustic environments (such as the cocktail party). To this end, we here assess the feasibility of using a novel, ultra-wearable ear-EEG device to classify the selective attention of normal-hearing listeners who participated in a two-talker competing-speakers experiment.

\textit{Approach.} Eighteen participants took part in a diotic listening task, whereby they were asked to attend to one narrator whilst ignoring the other. Encoding models were estimated from the recorded signals, and these confirmed that the device has the ability to capture auditory responses that are consistent with those reported in high-density EEG studies. Several state-of-the-art auditory attention decoding algorithms were next compared, including stimulus-reconstruction algorithms based on linear regression as well as non-linear deep neural networks, and canonical correlation analysis (CCA).

\textit{Main results.} Meaningful markers of selective auditory attention could be extracted from the ear-EEG signals of all 18 participants, even when those markers were derived from relatively short EEG segments of just \qty{5}{s} in duration. Algorithms which related the EEG signals to the rising edges of the speech temporal envelope (onset envelope) were more successful than those which made use of the temporal envelope itself. The CCA algorithm achieved the highest mean attention decoding accuracy, although differences between the performances of the three algorithms were both small and not statistically significant when EEG segments of short durations were employed.

\textit{Significance.} The results demonstrate that our ultra-wearable ear-EEG device offers promising prospects for wearable auditory monitoring. In the context of cognitively-steered hearing aids, the ear-EEG-derived attention markers provide indication of the focus of a listener's auditory attention; these could be fused with other signals available to a smart hearing device. Our results show that whilst the choice of attention decoding algorithm is not critical, the design of the speech feature (temporal envelope versus its onsets) is important. Therefore, future research should further pinpoint the features which are most appropriate for auditory attention decoding.

\end{abstract}

%
%
%
%
%

\section{Introduction}

People with hearing loss often struggle to comprehend speech in noisy environments, even when they wear their hearing aids~\cite{LesicaTINS2018, McCormackIJAud2013}. Various interventions have been developed to address this problem; for example, modern hearing aids incorporate advanced noise suppression technologies, which can reduce the impact of interfering sounds such as wind and other non-speech noises~\cite{BentlerTrendsAmplif2006}. In cocktail-party scenarios, where the interfering sounds are also speech, a hearing aid must first identify the target speaker, and then selectively amplify that voice alone. If the target speaker is known \textit{a priori} (for example, it is the teacher in a classroom environment), then the voice of that speaker can be wirelessly transmitted to hearing aid users via a hearing loop~\cite{YanzSemHear2003}. In more spontaneous settings such as social scenarios, beamforming technologies are usually utilised~\cite{DocloBookchapter2010}. These rely on directional microphones to identify the direction of the listener's head, and attempt to selectively amplify sounds emanating from that direction.

Amongst these interventions, only beamforming hearing aids offer users the ability to selectively and spontaneously choose which voice to enhance. In everyday environments, however, beamforming hearing aids often provide only limited benefits to their users~\cite{MagnussonIJAud2012, CordJAMAcadAud2004}. This limitation arises because individuals do not constantly orient their head towards a target speaker during naturalistic listening. Furthermore, interfering speech can easily divert the listener's attention and shift the direction of their gaze away from the intended speaker~\cite{RickettsJSLHR2008}. Whilst listeners can learn to adapt their behaviours in order to experience the maximum benefit from their beamforming hearing aids, even with practice they may struggle to quickly reorient their gaze when the focus of their attention changes location or is re-directed. This means that beamforming technology can actually increase the difficulty of locating off-axis sources for hearing-aid users~\cite{BrimijoinEarHear2014, ArcherBoydJHearRes2018}.

Alternative techniques for determining the focus of a listener's attention may be used in place of, or in conjunction with, beamforming algorithms. Recent research has demonstrated that the focus of a listener's selective attention to speech can be decoded from neuroimaging data~\cite{kidmose2010yarbus, Looney2010IJCNN, OSullivan2CerCor2014}. If a discrete and wearable brain monitoring device can be incorporated into a hearing aid, then it may be possible for the user to control the hearing aid through their attentional focus alone~\cite{GeirnaertSigMag2021, BleichnerJNE2016, FiedlerJNE2017}. Most studies identify electroencephalography (EEG) as the most suitable neuroimaging modality for real-world auditory attention decoding, due to its relative accessibility, its high temporal resolution, and the variety of wearable EEG devices which are already commercialised.

Usually, auditory attention is decoded from EEG recordings through a technique known as stimulus-reconstruction~\cite{OSullivan2CerCor2014, WongFrontNeuro2018, EtardNeuroImage2019}. This involves training regression models (known as backward models) to reconstruct features of the speech stimuli from time-aligned EEG signals. The feature of choice is typically taken to be the temporal envelope of speech, since this feature is strongly represented in EEG responses to speech~\cite{DingFrontHN2014}. Moreover, the strength of the EEG response to the speech envelope is modulated by selective attention, with responses to the attended speech envelope being dominant~\cite{Looney2010IJCNN}. Typically, the strength of the response is assessed by determining the accuracy with which the attended and ignored speech envelopes can be reconstructed from the EEG signals using regression models, by taking the Pearson correlation between the reconstructed envelopes and the ground-truth envelopes~\cite{OSullivan2CerCor2014}. The speech stream which yields the greatest Pearson correlation is then taken as the estimated attended stream.

Studies which make use of the stimulus-reconstruction approach to auditory attention decoding utilise underlying regression models which fall into two broad categories: linear models (typically based on ridge-regularised least-squares regression), and non-linear models implemented as deep neural networks~\cite{OSullivan2CerCor2014,CiccarelliSciRep2019,GeirnaertSigMag2021,puffay2023JNE,deTaillezEJNeuro2018,ThorntonJNE2022}. Stimulus-reconstruction approaches based on deep neural networks typically achieve higher attention decoding accuracies than their linear counterparts, since can account for the non-linear processing of the human auditory system. Some studies have also demonstrated that deep neural networks can generalise between participants and datasets remarkably well~\cite{Thornton2023OJSP, Accou2023SciRep}. As an alternative to the stimulus-reconstruction approach, canonical correlation analysis (CCA) can be used to relate EEG recordings to the speech envelope, and subsequently decode auditory attention~\cite{deCheveign2018NeuroImage, Hjortkjr2024BioArx}. This algorithm applies linear transformations to \textit{both} the EEG signals as well as the speech envelope signals in order to produce pairs of maximally-correlated components. By applying a linear classifier to the resulting correlation coefficients, a marker of selective attention to each competing speech stream can be obtained. Geirnaert~\textit{et al.}~\cite{GeirnaertSigMag2021} identified CCA as the best auditory attention decoding algorithm, outperforming all linear and non-linear stimulus-reconstruction approaches.

Most auditory attention decoding studies make use of EEG signals which were collected from multiple electrodes placed across the scalp of each participant. The EEG electrodes are typically fitted with a headcap and conductive gel in order to achieve a low and stable impedance across the skin-electrode interface. In contrast, the sensor montage employed by any prospective EEG-steered hearing aid must be much more dicrete and easy to set-up. Ideally, it will consist of just a few electrodes, and the device will not rely upon the application of conductive gel, which dries out with time (introducing noise) and can be difficult to apply properly for inexperienced users.

Currently, ear-centric sensor montages are popular choices for wearable EEG-monitoring devices~\cite{LooneyPulse2012, DebenerSciRep2015, WangNatCom2023, azemiPatent2023}. It is already established that such ear-centric devices can be used for long-term and discrete auditory monitoring, for example by detecting steady-state or subcortical responses to clicks and tones~\cite{Looney2011EMBS, Looney2014Chapter, kidmosePatent2015, Garrett2019FNINS}. In the context of auditory attention decoding, two types of ear-EEG devices have been studied: the concealed EEG (cEEG) device, which employs a C-shaped array of around-the-ear electrodes; and in-ear sensors, which sit entirely within the ear canal~\cite{BleichnerJNE2016, FiedlerJNE2017, HoltzeFrontNeuro2022}. In the study by Fiedler~\textit{et al.}~\cite{FiedlerJNE2017}, the authors found that auditory attention could not be decoded with significance when only in-ear electrodes were used, even in a binaural configuration. This result is to be expected, since EEG recorded from electrodes with a very small spatial separation has a very low signal-to-noise ratio (SNR)~\cite{MeiserBrainTop2020, YariciFnins2023}. Instead of using in-ear electrodes exclusively, Fiedler~\textit{et al.} proposed to reference the in-ear electrodes to a nearby location on the scalp; the FT7 location was identified as a promising candidate. The decoding accuracies achieved in that study were similar to those achieved using a binaural cEEG montage in the study by Holtze~\textit{et al.}~\cite{HoltzeFrontNeuro2022}. 

In the study of Fiedler~\textit{et al.}, the authors followed the usual stimulus-reconstruction approach to decode auditory attention to continuous speech in two-talkers conditions~\cite{FiedlerJNE2017}. However, in place of the commonly-used temporal envelope feature of speech, the authors made use of the onset envelope feature. The onset envelope is the defined as the half-wave rectified derivative of the temporal envelope, and therefore captures the rising edges of the temporal envelope which are predominantly driven by word and syllable onsets~\cite{HertrichPschoPhysiol2011}.

In this work, we demonstrate how a dry-contact, wirelessly connected ear-EEG device can be used to decode the focus of a listener's auditory attention in two-talker competing-speakers scenarios. The sensor configuration is similar to that proposed by Fiedler~\textit{et al.} in that it consists of two in-ear electrodes, and one adjacent scalp electrode. We adopt three attention decoding algorithms: stimulus reconstruction based on linear models and deep neural networks, and CCA. We further provide a comparison of the attention decoding performance when the temporal envelope feature is used, versus when its onsets is used.

\section{Materials and methods}

\subsection{Subjects and stimuli}

\begin{figure}
    \centering
    \includegraphics[width=1.\textwidth]{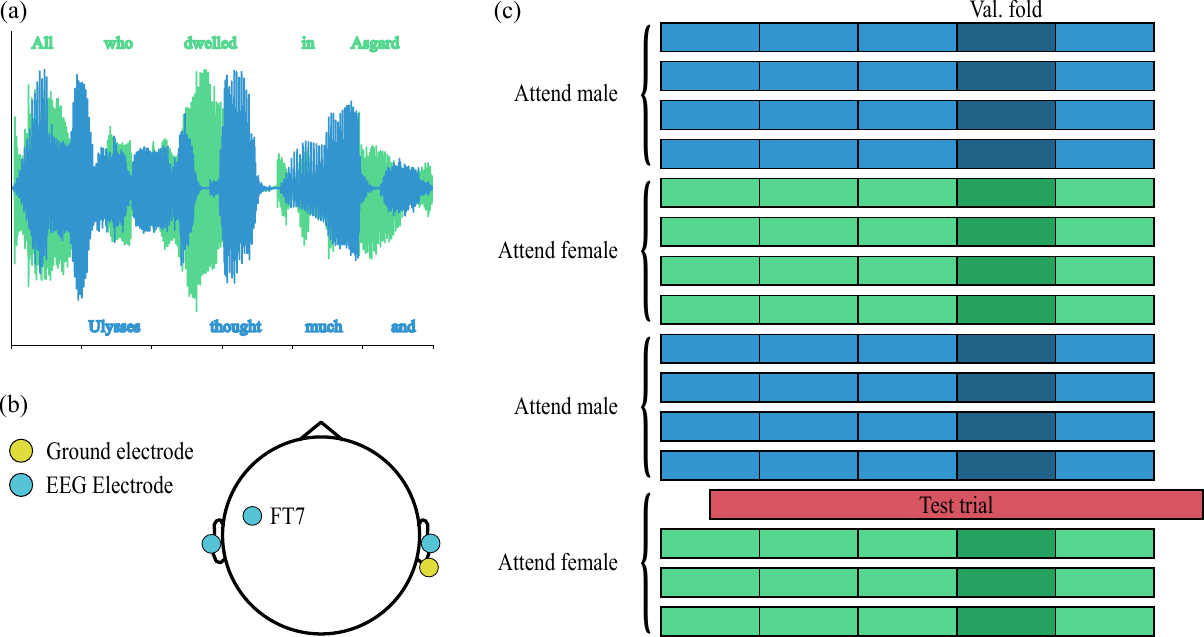}
   \caption{Overview of the experiment and analysis protocols. a) Participants listened to a male voice and a female voice which narrated two distinct audiobooks at the same time. They were asked to direct their attention to one of the voices, and to ignore the other. b) The dry-contact EEG sensor montage employed in this work consisted of  two in-ear electrodes which were referenced to the FT7 scalp channel, and a clip-on ground electrode which was attached to the right earlobe. c) Participants listened to four short stories, each of which was split into four trials of approximately \qty{2.5}{minutes} in length. The to-be attended speaker alternated with each story. A nested-cross-validation procedure was used to train and test the attention decoders: data from each trial was held-out in turn; the remaining data were split into five folds. Four of those folds were used to train attention decoders with different hyperparameters, and the remaining fold was used to select the best hyperparameter configurations.}
    \label{fig:overview}
\end{figure}

Eighteen young, normal-hearing participants (median age: 23 years; 10 males, 8 females) took part in a selective auditory attention experiment. The experimental protocol was approved by the Imperial College Research Ethics Committee (approval number 19IC5388 No A1; approved 20\textsuperscript{th} September 2019). The experiment consisted of 16 trials: during each trial, two audiobooks chapters narrated by a male and a female speaker were presented simultaneously to the participant at a sampling frequency of \qty{44.1}{kHz} via Seinheiser HD450 headphones, which were placed over the form factors of the in-ear electrodes. The participant was instructed to attend to either the male speaker or the female speaker, and to ignore the other speaker. On average, each trial lasted for 150 seconds.

The speech material was narrated by the same two speakers for all 16 trials. For each consecutive four trials, the participants were directed to attend to the same speaker, who narrated four consecutive chapters of an audiobook. After each set of four trials, the participants were directed to switch their attention to the other speaker for the next four trials. The speech material was presented diotically (i.e. the narrators were not spatially separated) at a sound pressure level of 75 dB SPL (measured via a Bruel \& Kjaer type-4157 ear simulator.)

\subsection{EEG data acquisition}

During the experiment, the electroencephalograms of the participants was recorded from an ultra-wearable, wirelessly-enabled two-channel ear-EEG device. Our in-house, dry-contact electrodes were inserted into the ear canals of the participants, and a third reference electrode was placed on the left lower-forehead (corresponding to the FT7 location in the standard 10-20 system). A ground electrode was attached to the right earlobe. This recording system therefore acquires EEG from a bilateral (cross-head) channel and a unilateral channel. Accelerometers and microphones are built into the form factors of the insert earphones: signals acquired from these sensors can be used for the purpose of denoising electrophysiological signals, as demonstrated in previous work from our research group~\cite{OcchipintiIJCNN2022, HammourEMBC2021}. The device acquires all sensor signals at a sampling frequency of 256 Hz. During the experiment, the mixed audio of the two speech streams was acquired through the in-built microphone. During offline processing, the recorded audio signal was used to align the original audio signals with the recorded EEG signals via a cross-correlation-based method.

\subsection{Signal pre-processing}

We related the EEG recordings to the temporal envelopes and the onset envelopes of each speaker separately. There are several types of temporal envelope which have been explored in the literature; we used the auditory-inspired envelope first proposed by Biesmans~\textit{et al.}~\cite{BiesmansTransNeurRehab2017}. This envelope is calculated by filtering the audio signals into 28 sub-bands via a gammatone filterbank. The centre frequencies of the filterbank are spaced equidistantly between \qty{50}{Hz} and \qty{5}{kHz} on an equivalent rectangular bandwidth scale. The signal in each sub-band is then half-wave rectified, and the resulting signals are averaged together to form a stimulus envelope. As in the study by Fielder~\textit{et al.}~\cite{FiedlerJNE2017}, the onset envelope feature was formed by half-wave rectifying the first derivative of the temporal envelope feature.

The EEG signals were preprocessed by first de-trending them via a high-pass filter with a cutoff frequency of \qty{0.5}{Hz} (Type-2 Hamming-sinc FIR filter of order 1691 with a \qty{-6}{dB} stopband attenuation at \qty{0.25}{Hz}). The EEG and envelope signals were finally resampled to a sampling frequency of \qty{64}{Hz} and were  standardised by removing the mean of each EEG channel respectively of the envelope signal and by dividing each time series by its standard deviation.

\subsection{Forward models}

Forward models, known also as temporal response functions (TRFs), are estimates of the impulse response functions which relate the EEG signals recorded from each channel to a particular feature present in the speech stimulus~\cite{Lalor2010}. The features that we considered in this case were the temporal envelope of speech as well as its onsets.

In practice, TRFs can be estimated by fitting FIR filters to predict the EEG signals from the speech-feature signals, with the filter coefficients themselves representing the TRF estimates. We used 160 filter taps, with each corresponding to a response latency ranging from \qty{-1}{s} (pre-stimulus) to \qty{+1.5}{s} (post-stimulus). Following Biesmans~\textit{et al.}~\cite{Biesmans2017}, the ridge regularisation parameter of each TRF was selected as the mean eigenvalue of the speech-feature autocovariance matrix. The TRFs were estimated using a cross-validation procedure: for each participant, data from one of the 16 trials was held-out for evaluation. A TRF was then estimated using the data from the remaining 15 trials. Overall, 16 TRFs were fitted per-participant using this procedure, which were averaged in order to produce a single TRF per participant. In this way, we fitted distinct TRFs for each EEG channel, each stimulus feature (the speech envelope as well as its onsets), and each speaker (the attended speaker and the ignored speaker). In order to assess whether the TRFs were influenced by selective attention, we also formed `difference TRFs' by subtracting the ignored-speaker TRFs from the attended-speaker TRFs.

It is important to make sure that the TRFs are meaningful before interpreting their coefficients. To this end, a single-sample permutation-based cluster test was used to test whether the instantaneous power of the TRF coefficients was statistically different to random chance. First, a set of null-TRFs was constructed by repeated the cross-validation feature described above, but with temporally misaligned EEG signals and speech-feature signals. We repeated this procedure using 500 different temporal misalignments, yielding 500 null-TRFs per participant, per held-out trial. Temporal clusters in the TRFs were formed by marking samples at which the TRFs' instantaneous power exceeded the 99\textsuperscript{th} percentile of the instantaneous power distribution obtained from the null-TRFs. The size of the largest cluster formed the test statistic of the permutation-based cluster analysis. A null distribution for this test statistic was formed by randomly permuting the signs of the participant's TRFs prior to averaging them; we used 1000 separate permutations to form this distribution. Although the cluster-based permutation tests only test the null hypothesis that each TRF as a whole is different to random chance, clusters with low p-values can still be visualised and interpreted: we retained and inspected all clusters with a p-value less than $p=\frac{0.05}{12}$ (i.e. we Bonferroni corrected for the $12$ TRFs shown in Figure~\ref{fig:TRFs}.)

\subsection{Linear backward models}

Like forward models, linear backward models are FIR filters which relate the EEG recordings to features of the auditory stimuli (in this case, the envelopes and onset envelopes of the competing speech streams). However, unlike forward models, backward models predict the speech features from the EEG recordings. The same algorithm (regularised ridge-regression) was used to fit the backward models: in this case, both EEG channels were used to predict speech features, and we considered 64 filter taps corresponding to latencies between \SIrange{0}{1}{s}. In other words, EEG samples which were delayed up to \qty{1}{s} relative to the audio stimulus were used to predict the speech features.

We computed participant-specific linear backward models and tuned the ridge regularisation parameter using the nested cross-validation procedure outlined in Figure~\ref{fig:overview}. Data from each trial was held-out in turn for the purpose of evaluating the final models. The remaining data was split into five equally-lengthed folds; each fold was held-out in turn, and the remaining four folds were used to train 19 models with regularisation parameters spaced equally on a log scale between $10^{-9}$ and $10^{9}$. The model which achieved the highest correlation coefficient on the held-out fold was submitted for final evaluation on the held-out trial. As with the forward models, we fitted backward models for each speech feature and each competing speaker. A null distribution for the correlation coefficients was constructed by correlating the reconstructed speech features from a particular testing trial against the actual speech features taken from other trials.

\subsection{CNN-based backward models}

The CNN-based backward models were direct analogues of the linear backward models in the sense that they used a temporal window of 2-channel EEG recordings, of \qty{1}{s} in duration, to reconstruct the speech features which occurred at the onset of the temporal window. However, as universal function approximators, the CNN-based backward models implement non-linear mappings between the EEG recordings and the speech-feature signals, unlike the linear backward models~\cite{Mandic2001Book}. This means that they may be better placed to account for the fundamentally non-linear nature of the human auditory system~\cite{deTaillezEJNeuro2018}.

The CNN architecture used in this work consists of several convolutional blocks which are linked together in a feed-forward manner with skip connections. Each convolutional block consists of a convolutional layer, followed by a rectified linear unit (ReLU) activation function, a batch normalisation (BatchNorm) layer, and an average pooling layer. Convolutional layers of artificial neurons act as pattern detectors, since they implement a set of learned multi-channel matched filters; by sequentially stacking convolutional layers, CNNs can detect complex hierarchical patterns~\cite{Mandic2023MFSysManCyb}. Non-linear activation functions such as the ReLU are essential elements of artificial neural networks which allow them to approximate non-linear functions~\cite{Mandic2001Book, Schmidhuber2015NNs}. BatchNorm performs a learned normalisation of the outputs of the convolutional layer, and has been reported to lead to faster and more stable training of CNNs~\cite{Ioffe2015ICMLBN, Santurkar2018NeuripsWhyBN}. Average pooling layers perform downsampling on their inputs through averaging adjacent temporal samples. Since some studies in EEG decoding have reported improved performance when skip connections are used, we utilised them here~\cite{Accou2023SciRep, He2016CVPR}. After the last convolutional block, a linear combination of the transformed EEG signals is taken to produce an estimate of the envelope (or its onsets) of the attended speech stream.

Participant-specific CNNs were trained using the same nested cross-validation scheme as the linear backward models (see Figure~\ref{fig:overview}). The parameters of the CNNs were tuned using the Adam optimiser with a learning rate of $0.001$~\cite{Kingma2015ICLR}. Using the inner cross-validation loop, we tuned two hyperparameters: the kernel size of the convolutional layers (3 or 5) and the number of convolutional blocks (1, 2 or 3).

\subsection{Auditory attention decoding via backward models}
\label{sec:aad-bm}

The following procedure was used to perform auditory attention decoding using both types of backward models. Data from the held-out trials were first split into temporal segments of lengths ranging from \qty{0.1}{s} to \qty{30}{s}. There was a \qty{1}{s} hop length between adjacent segments. For each segment, we obtained reconstructed speech features from the attended-speaker backward models, and these were correlated against the corresponding features derived from both the attended and ignored speech streams. The difference in correlation coefficients, $\Delta\rho = \rho_\textrm{attended} - \rho_\textrm{ignored}$, served as a marker of selective auditory attention. A null distribution for the attention markers was obtained by correlating the reconstructed speech features from each particular segment with the actual speech-features corresponding to a different, randomly selected segment. Single-tailed, unpaired t-tests were used to compare the attention markers against the null markers for each participant.

\subsection{Canonical correlation analysis}

Canonical correlation analysis applies linear filters to both the EEG as well as the speech-feature signals, in order to yield new signals, or `canonical components', which are maximally correlated. Because of this, CCA can be considered as the simultaneous application of forward and backward models. The CCA algorithm yields multiple pairs of canonical components, of which the first pair are the most highly correlated, the second pair are the next most highly correlated, and so on. Importantly, different pairs of canonical components are mutually uncorrelated, i.e. neither of the first pair of components are correlated with either of the second, \emph{et cetera}.

In Geirnaert~\textit{et al.}~\cite{GeirnaertSigMag2021}, the authors used CCA to perform auditory attention decoding by performing the following steps: first, they trained the CCA algorithm to maximise the correlation between the EEG signals and the speech envelope of the attended speaker. Then, they passed the EEG signals and corresponding attended and ignored speech envelopes through the algorithm, obtaining two vectors of correlation coefficients $\mathbf{\rho_\textrm{attended}}$ and $\mathbf{\rho_\textrm{ignored}}$; these correspond to the attended and ignored speech streams respectively. Finally, they trained a linear classifier (linear discriminant analysis, LDA) to distinguish between the correlation difference vectors $\mathbf{\rho_\textrm{attended}} - \mathbf{\rho_\textrm{ignored}}$ and $\mathbf{\rho_\textrm{ignored}} - \mathbf{\rho_\textrm{attended}}$, thus estimating the identity of the attended speaker.
Here, we trained CCA-based attention decoders in same fashion as Geirnaert~\textit{et al.} using the nested cross-validation procedure described above. We trained the LDA classifiers on correlation difference vectors derived from short temporal speech/EEG segments of \qty{5}{s} in duration. As with the backward models, we evaluated the CCA-based algorithms using various segment durations with a \qty{1}{s} hop length between segments. The number of canonical components to consider (i.e. the dimensionality of the correlation difference vectors) was tuned via the inner cross-validation loop. Following Geirnaert~\textit{et al.}, we did not perform any dimensionality reduction on the original signals prior to applying the CCA algorithm.

\section{Results}

\subsection{Forward models}

\begin{figure}
    \centering
    \includegraphics[width=1.\textwidth]{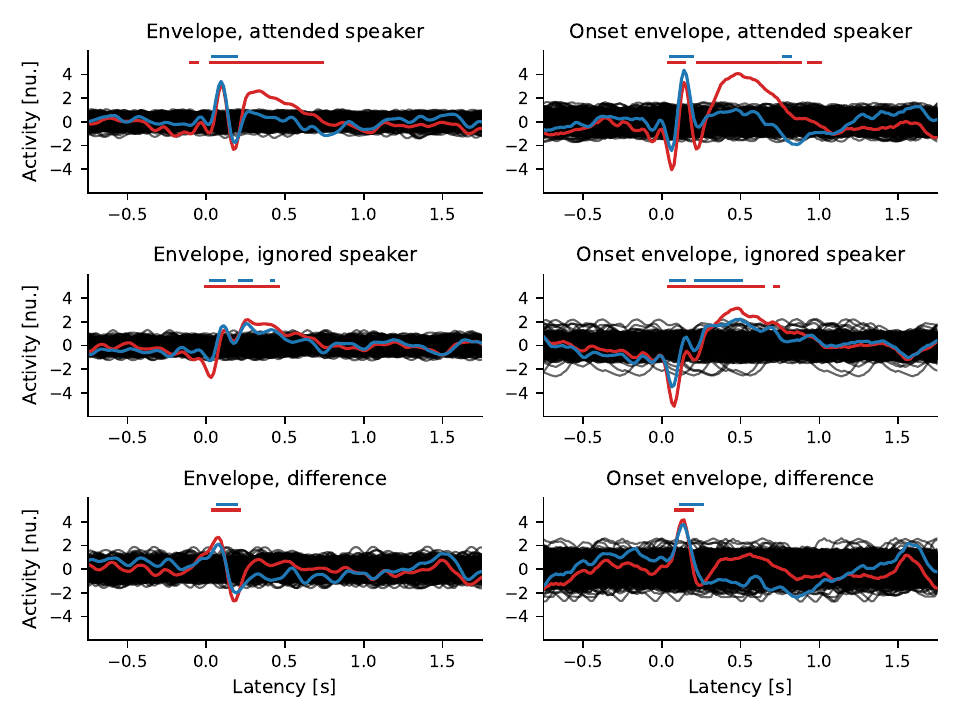}
   \caption{Temporal response functions for both the uni-lateral (blue) and bi-lateral (red) EEG channels. The first column shows TRFs computed from the temporal speech envelopes, and the second the onset-envelope TRFs. The first two rows show the grand-average TRFs for attended and ignored speech, respectively, and the bottom row shows the difference between the attended and ignored TRFs. Black lines represent null-TRFs, which were computed after temporally misaligning the EEG and speech-feature signals. Solid red and blue bars indicate temporal clusters with p-values smaller than $0.05/12$, as calculated through a cluster-based permutation test.}
    \label{fig:TRFs}
\end{figure}

Temporal response functions were fitted to relate the EEG recordings to the speech envelopes and their onsets. The TRFs were estimated for both the attended speaker and the ignored speaker. In order to assess whether the TRFs were modulated by selective attention, we formed difference TRFs by subtracting the ignored speech TRFs from the attended speech TRFs.

Single-sample cluster-based permutation tests revealed that all of the TRFs and difference TRFs yielded patterns of instantaneous power which were significantly different to random chance. 
The temporal response functions for the attended and ignored speakers are depicted in the first two rows of Figure~\ref{fig:TRFs}. They exhibit morphologies which are typical of TRFs obtained in higher-density EEG studies, with clear and significant components around latencies of \qty{100}{ms} and \qty{200}{ms}. The onset envelope TRFs also show pronounced activity at a much later latency of around \qty{500}{ms}. The bottom row of Figure~\ref{fig:TRFs} shows the difference TRFs. For the envelope feature, the amplitude of the first two peaks of the unilateral-channel TRF are significantly modulated by selective attention. As for the onset envelope feature, the first positive peak of both the unilateral and bilateral channel TRFs is significantly modulated by selective attention. In contrast, there is no significant attentional modulation of the later components.

\subsection{Linear backward models}

\begin{figure}
    \centering
    \subfigure[]{\includegraphics[width=.49\textwidth]{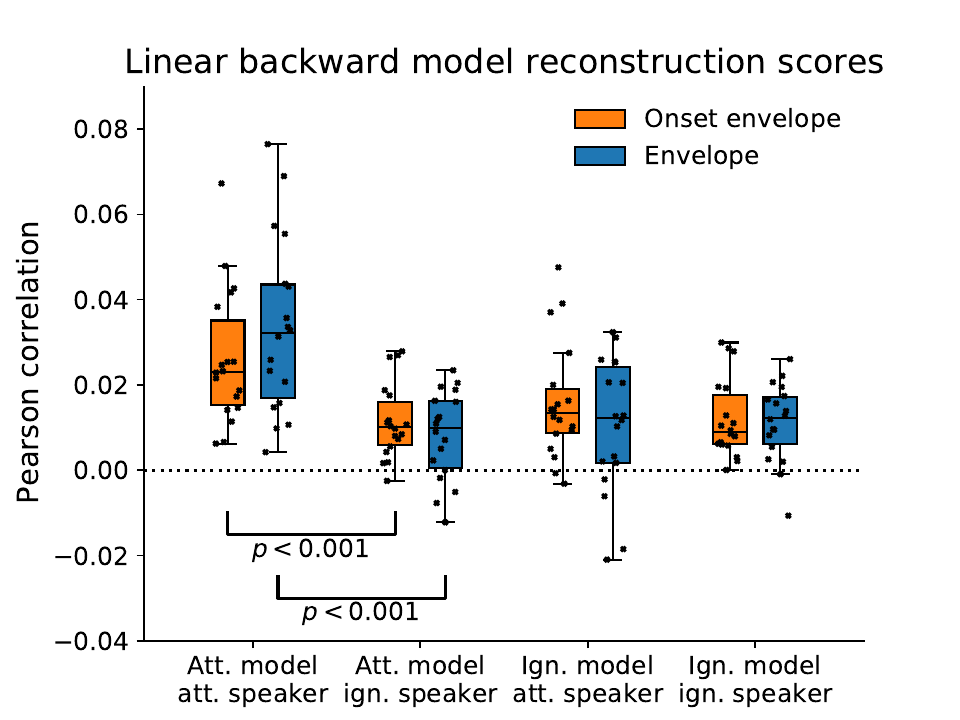}}
    \subfigure[]{\includegraphics[width=.49\textwidth]{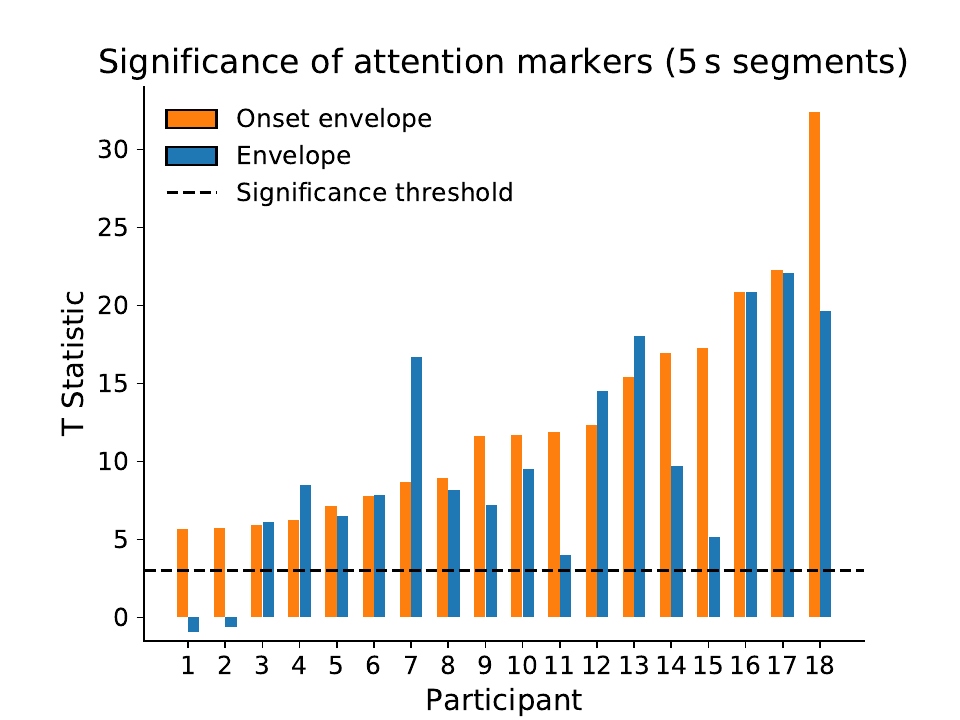}}
   \caption{Performance of the linear backward models. a) The reconstructed speech features produced by the models were correlated against the actual features of both the attended and ignored speech streams. Correlation coefficients were calculated for each testing trial and participant, and the average correlation coefficients for each participant are displayed as black markers. b) Attention markers, defined as the difference in the correlation coefficients obtained for the attended speech stream and the ignored speech stream, were calculated for each participant. The correlation coefficients were computed using short data segments of \qty{5}{s} in duration. Unpaired t-tests were used to compare the attention markers to appropriate null distributions, and the resulting t-statistics are displayed in the plot. A dashed black line shows the Bonferroni-corrected significance threshold for single-tailed tests.}
    \label{fig:backward-models}
\end{figure}

Linear backward models were fitted to predict features of the attended and ignored speech streams from the two-channel EEG signals. Correlation coefficients for each held-out trial were averaged for each participant, and the results are shown in Figure~\ref{fig:backward-models}a. All eight groups of correlation coefficients have means which are statistically greater than their corresponding null distribution (all $p\ll 0.0001$; single-tailed unpaired t-tests with Bonferroni correction). The models which were trained using features from the attended speech stream produced reconstructed features which were more highly correlated with features of the attended speech stream than those of the ignored speech stream ($p\ll 0.0001$ for both the envelope feature, and its onsets; single-tailed paired t-tests.)

To further investigate how the attended-speaker models could be used for practical auditory attention decoding, we divided each held-out trial into short temporal segments of \qty{5}{s} in duration (with no overlap). Markers of selective auditory attention were derived from these segments using the procedure outlined in Section~\ref{sec:aad-bm}. To see if the attention markers obtained from such short segments were meaningful for individual participants, we performed single-tailed unpaired t-tests to compare them to their null distributions; the resulting t-statistics are depicted in Figure~\ref{fig:backward-models}b. Backward models which used the onset envelope feature yielded significant attention markers for all 18 participants, whereas the backward models based on the envelope feature yielded significant attention markers for all but two participants.

\subsection{Comparison of decoding algorithms}

\begin{figure}
    \centering
    \includegraphics[width=\textwidth]{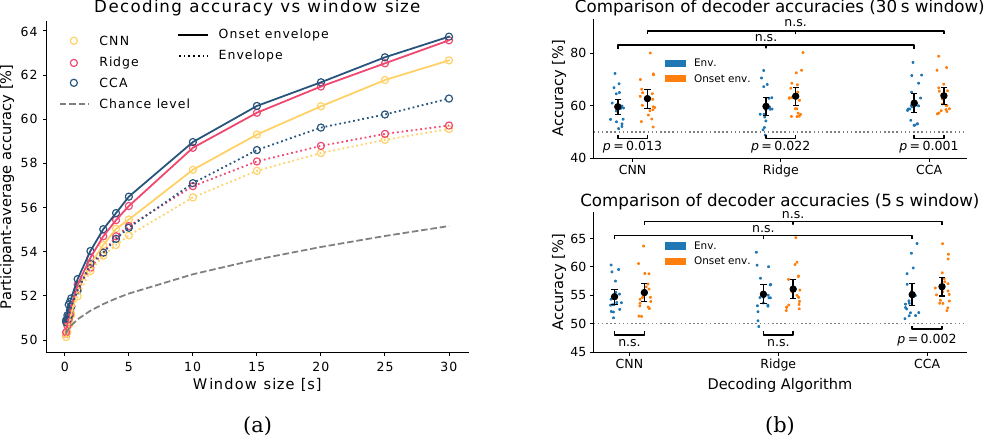}
   \caption{Performance of the attention decoding algorithms. a) The decoding accuracies for individual participants were averaged and plotted against segment length. The results for the algorithms which used the speech envelope feature are displayed with broken lines, and solid lines are used for those which utilised the onset envelope. The grey dashed line represents the upper limit of the 95\% confidence interval of a random binary classifier. b) The attention decoding accuracies were broken down for individual participants for a longer segment length of \qty{30}{s} (top) and a shorter window length of \qty{5}{s} (bottom), and the mean attention decoding accuracies of different algorithms were assessed for statistical differences using single-tailed, paired t-tests.}
    \label{fig:acc-v-ws}
\end{figure}

The three attention decoding algorithms were evaluated using segment lengths ranging from \qty{0.1}{s} to \qty{30}{s}. For each algorithm and speech feature, the mean attention decoding accuracy is plotted against segment length in Figure~\ref{fig:acc-v-ws}a. Overall, the mean accuracies lie above the chance level (defined as the $95^{\textrm{th}}$ percentile of a random binary classifier) across a wide range of window sizes. The decoding algorithms which utilised the onset envelopes achieved higher mean decoding accuracies than those based on the envelopes themselves, and in general the CCA decoders achieved the highest mean decoding accuracies, and the CNN-based decoders achieved the lowest. We tested whether these differences were statistically significant using single-tailed paired t-tests, for both a short segment length of \qty{5}{s} as well as a longer segment length of \qty{30}{s}. The results are shown in Figure~\ref{fig:acc-v-ws}. For the \qty{5}{s} segment length, we did not find a significant difference between mean accuracies of any of the algorithms, except for the two CCA-based decoders which were based on either the speech envelope or its onsets. For the longer window length of \qty{30}{s}, however, significant differences emerged between the envelope-based decoders and the onset-envelope-based decoders for all three decoder types. We were nonetheless still unable to detect statistically significant differences between the mean accuracies of different decoder types.

\section{Discussion and conclusion}

We have demonstrated that our novel, ultra-wearable ear-EEG device captures auditory responses to the envelope of speech (and its onsets) from normal-hearing human listeners. The recorded responses were shown to be modulated by the focus of the listeners’ selective attention to speech. We were therefore able to decode which of the two speakers the participants were attending to, using a variety of auditory attention decoding algorithms.

Temporal response functions obtained from the ear-EEG device exhibit a clear morphology which is consistent with the existing scalp-EEG-based literature, with strong components at latencies of around 100-200 ms. The onset-envelope TRFs also exhibit broad regions of activity at a much later latency of around 500 ms which was not reported in the study by Fielder~\textit{et al.}~\cite{FiedlerJNE2017}. This is because the high-pass filter which we pre-processed the EEG signals with had a much lower cutoff frequency than that which was used in that study (\qty{0.5}{Hz} vs \qty{2}{Hz}). This means that our TRFs are influenced by contributions from very low-frequency cortical tracking of words and syllables.

Linear backward models were trained to relate the two-channel EEG recordings to speech features of the attended speaker and the ignored speaker. For both speech features, the attended-speaker models better predicted the features of the attended speech stream than those of the ignored speech stream. By contrast, the ignored-speaker models achieved similar correlation scores for both speakers. We therefore used the attended-speaker models as auditory attention decoders. Markers of selective auditory attention were formed by subtracting the correlation coefficients obtained for the ignored speech stream from those obtained for the attended speech stream. We found that the onset-envelope-based attention decoders produced statistically meaningful attention markers for all 18 participants when a short, practical segment length of \qty{5}{s} was used. For the envelope-based attention decoder, the attention markers were statistically meaningful for 16 of the 18 participants. Overall, we observed that the test statistics for the onset-envelope-based decoders were greater than those for the envelope-based decoders for two-thirds of the participants.

Three types of algorithms were used to perform auditory attention decoding: the linear backward models; non-linear CNN-based backward models; and CCA-based decoders. Overall, the attention decoding accuracies were not that high, lying in the region of 55\% for short segment lengths of \qty{5}{s} in duration. Clearly, devices such as smart hearing aids which rely on EEG-based auditory attention decoding are unlikely to provide any benefit to the user unless the attention decoding accuracy could be significantly enhanced. One possibility would be to employ state-space models to obtain less noisy estimates of the attended speaker label from the original attention markers: Akram~\textit{et al.}~\cite{Akram2014NIPS,Miran2018FontNeuro} developed an algorithm which, at its core, employs a Kalman filter to denoise the correlation-based attention markers, leading to a moderate improvement in attention decoding accuracy; Hjortkjaer~\textit{et al.}~\cite{Hjortkjaer2024preprint} proposed an alternative state-space model which weights the importance of incoming attention markers based on the recent performance of the underlying attention decoder; many more possibilities exist. An orthogonal approach to improving the attention decoding accuracy of a hearing aid system, for example, could be by leveraging the information available in other sensing modalities via sensor fusion. For example, additional markers of selective attention can be derived from eye-tracking and electro-oculography signals~\cite{Skoglund2022FrontNeuro, Jin2018NatCom}.

We did not detect any significant difference in mean accuracy between the three types of decoding algorithm for short segment lengths of \qty{5}{s} in duration, nor for longer segment lengths of \qty{30}{s}. It is possible that with only 18 participants, we did not have enough statistical power to detect such a difference; nevertheless, it is clear that there is no practically significant difference in the performance of the three decoding algorithms, since the mean accuracy only varies by a few percentage points between them. The deep-learning-based decoders achieved lower mean decoding accuracies than their linear counterparts for all segment lengths, which contrasts with the recent scalp-EEG literature~\cite{deTaillezEJNeuro2018,ThorntonJNE2022}. This could be due to the that fact that the ear-EEG signals offered a poorer SNR, since the dry-contact electrodes were placed in locations which are prone to mastication artifacts; the CNNs may therefore have overfitted to artifactual signals during training.

The choice of speech feature (temporal envelope versus its onsets) did  impact  the final attention decoding accuracy. These speech features were based on biologically-inspired heuristics, and it is possible that by refining them (and possibly tailoring them to individuals, since Figure~\ref{fig:backward-models} shows that sometimes one feature is better than another for some participants), further improvements in auditory attention decoding may be made. Such an approach could employ techniques of deep learning to approximate the representations of speech which are present in ear-EEG recordings, following similar approaches to modelling invasively-measured electrophysiological responses to speech~\cite{Drakopoulos2024preprint}.

\section*{Data availability statement}

The original ear-EEG recordings and speech material which were used in this work are available at \href{https://doi.org/10.5281/zenodo.10260082}{https://doi.org/10.5281/zenodo.10260082}.

\section*{Acknowledgements}

Michael Thornton was supported by the UKRI CDT in AI for Healthcare \href{http://ai4health.io}{http://ai4health.io} (Grant No. P/S023283/1).

\printbibliography[title={References}]
\end{document}